\documentclass[11pt]{article}
\usepackage{amssymb}
\usepackage{latexsym}
\usepackage{amsmath}
\usepackage{graphicx}
\usepackage{listings}
\usepackage{cite}

\def\al{\alpha}
\def\be{\beta}
\def\g{\gamma}
\def\L{\Lambda}
\def\l{\lambda}
\def\bq{\bar{q}}
\def\k{\kappa}
\def\ve{\varepsilon}
\title{\LARGE{Static spherically symmetric black hole's solution in Einstein-Maxwell-Yang-Mills-dilaton theory}}
\author{M. M. Stetsko\footnote{E-mail: mstetsko@gmail.com}\
\\
  {\small Department for Theoretical Physics, Ivan Franko National University of Lviv,}\\
{\small 12 Drahomanov Str., Lviv, UA-79005, Ukraine
         }}
\begin{document}
\maketitle

{\abstract{In this work a static spherically symmetric black hole's solution within the Einstein-Maxwell-Yang-Mills-dilaton theory is derived. The obtained solution is examined, in particular, we have calculated the Kretschmann scalar which allowed us to characterize singularity points. Using the obtained black hole's solution we have studied its thermodynamics. Namely, the temperature was calculated, the first law was written, and the heat capacity was studied. The extended thermodynamics approach is utilized to obtain the equation of state. Within the extended thermodynamics the Gibbs free energy is   derived and investigated. The analysis of the Gibbs free energy shows that below the critical temperature besides the first order phase transition in addition the zeroth order phase transition occurs, what is typical for other types of black holes with dilaton fields. Finally, we have shown that the critical exponents take the same values as for other types of the dilaton black holes.}}

\section{Introduction}
General Relativity is known to be the most successful theory of gravitation which explains extremely wide range of phenomena on vast space and time scales, namely it gives correct description of the planetary motion as well as it expounds the evolution of the Universe \cite{Will_LRR2014,Berti_CQG2015}. General Relativity has passed numerous experimental verifications with flying colours \cite{Abbott_PRL16}. Despite the persuasive success of General Relativity there are still some open issues, for instance we remark the Dark Matter/Dark Energy issue or the Inflation problem. To solve the mentioned above problems different approaches were developed and applied, some  of them were more revolutionary, where the basic principles of the theory were modified or extended, and the other ones were more conservative, where for instance additional material fields were included without serious modification of the basic principles \cite{Capo_PRp11,Clifton_PRp12,Heisenberg}. 

Black holes are known to be one of the most interesting objects of examination in General Relativity and in other theories of Gravity. This interest that firstly was motivated by Astrophysics and Cosmology now is stimulated by even a bit distant areas of research such as String Theory and the following from it AdS/CFT correspondence \cite{Maldacena_ATMP98,Witten_ATMP98}, which nowadays has numerous applications in different branches of Physics. The latter as it is known allows to develop some correspondence between Gravity in the bulk and Conformal Field Theory on the boundary, but it should also be noted, that now this correspondence is treated in a more general context. Thus the special interest is paid to the black holes in AdS space or in other spaces which are not asymptotically flat at the infinity. Among the others, the special attention is drawn to the black holes with dilaton fields. The dilaton fields as well as some other fields such as  for instance axions have their roots in the String Theory, in particular, they can be derived by virtue of the dimensional reduction procedure. The black holes' solutions with dilaton fields were firstly considered in late 80-ies and early 90-ies of the last century \cite{Gibbons_NPB88,Garfinkle_PRD91,Witten_PRD91}. Then they got broad attention among the scientific community and lots of new black holes were obtained and different aspects of their behaviour were studied \cite{Kallosh_PRD92,Gregory_PRD93,Rakhmanov_PRD94,Poletti_PRD94,
Gibbons_CQG95,Chan_NPB95,Cai_PRD96, Gao_PRD04,Yazadjiev_CQG05,Astefanesei_PRD06, Mann_JHEP06,Kunz_PLB06,Brihaye_CQG07,Charmousis_PRD09,Sheykhi_PRD07,
Sheykhi_PLB08,Sheykhi_PRD14,Fernando_PRD09,Kord_PRD15,
Hendi_PRD15,Dehyadegari_PRD17,Dayyani_2017,Pedraza_CQG19,Bravo_GaetePRD18}. A lot of these works take into account additional Maxwell field, it allows to obtain a kind of generalization of the well-known Reissner-Nordstrom solution and other charged black holes. Dilaton black holes with additional Maxwell field is the subject of deep interest, since their possible applications to various problems in different areas \cite{Goldstein_JHEP10,Arefeva_JHEP16}.

Nonabelian gauge fields can be treated as a some sort of generalization of the abelian ones. In spite of relatively long history of their investigation, they have not got so wide attention as their abelian counterparts, this fact can be explained by different factors, in particular by more complicated structure of the obtained solutions. For the first time a black hole's solution with nonabelian field was derived and studied by Yasskin in the middle of the 70-ies of the last century \cite{Yasskin_PRD75}, but his work did not attract much attention at that time. More than a decade later the interest to the black holes with nonabelian field was renewed \cite{Bartnik_PRL88,Bizon_PRL90,Volkov_JETP89,Kuenzle_JMP90}, this enthusiasm was mainly motivated by the development of the String Theory, which brought the interest to different kinds of fields and in particular to the nonabelian ones. The solutions with the asymptotically flat background were shown to be unstable \cite{Straumann_PLB90}, but it drew more extensive attention to the black holes' solutions with nonabelian fields \cite{Torii_PRD95,Volkov_PRD96,Volkov_PRep98,Mavromatos_JMP98}. Here we point out that  in addition to ``gravitational'' instability  investigated in \cite{Straumann_PLB90} additional mechanism of instability called as ``sphaleronic'' was also observed and examined \cite{Volkov_PLB95}. It was shown that the black holes was stable in the anti-de Sitter space \cite{Winstanley_CQG99,Bjoraker_PRL00,Bij_PLB02}. Colored black holes and Bartnik-McKinnon solution initially obtained \cite{Bizon_PRL90,Volkov_JETP89,Kuenzle_JMP90} and \cite{Bartnik_PRL88} respectively  were generalized with taking into account higher order curvature terms \cite{Brihaye_PLB03}, Yang-Mills black hole with dilaton field was obtained and investigated \cite{Lavrelashvili_NPB93,Radu_CQG05}. Higher dimensional black holes with higher order Yang-Mills hierarchy was obtained \cite{Brihaye_PRD07}. Black hole with superconducting horizon was derived \cite{Manvelyan_PLB09}. Black holes' and solitonic solutions in case of higher order gauge group and in more extended frameworks than the General Relativity were studied \cite{Baxter_PRD07,Lerida_PRD09,Ghosh_PLB11,Mann_PRD06,Cvetic_PRD10}.    It should be noted that the so-called Wu-Yang ansatz was extremely useful for obtaining black holes' solutions \cite{Mazhari_PRD07,Mazhari_PRD08,Mazhari_PLB08,Mazhari_GRG10,
Mazhari_PRD11,Mazhari_EPJC13}. Namely, it was used to obtain the solutions in case the nonabelian Lagrangian is of Born-Infeld type in \cite{Mazhari_PRD08,Mazhari_GRG10}, Lovelock and $F(R)$ gravity \cite{Mazhari_PLB08,Mazhari_PRD11,Mazhari_EPJC13}. The Wu-Yang ansatz was also used in case of topological black holes \cite{Bostani_MPLA10,Dehghani_IJMPD10}, black holes in nonminimal Einstein-Yang-Mills theory \cite{Balakin_PRD16}. Charged Einstein-Yang-Mills black hole within the gravity's rainbow was studied. \cite{Hendi_PLB18}. Black hole's solutions were derived and their various properties were investigated in massive \cite{Hendi_JHEP19} and dimensionally continued \cite{Ali_PRD19} gravities.

In our work we examine a static spherically symmetric black hole within the Einstein-Maxwell-Yang-Mills-dilaton theory. This paper can be considered as a generalization of our previous results in the framework  of Einstein-Maxwell-dilaton theory \cite{Stetsko_EYM20}. In some sense it can be treated also as a generalization of Einstein-Maxwell-dilaton theory \cite{Stetsko_EPJC19}. This work is organized as follows. In the second section we write the equations of motion in the Einstein-Maxwell-Yang-Mills-dilaton theory, obtain the static spherically symmetric solution and investigate it. In the third section we study the thermodynamics, namely we derive and examine the temperature, we write the first law of black hole thermodynamics, we also obtain the heat capacity which is important to characterize thermal stability. In the fourth part using the extended thermodynamics framework we obtain and investigate the equation of state, then we derive and study the Gibbs free energy. Finally, the fifth section contains some conclusions.

\section{Field equations for Einstein-Maxwell-Yang-Mills-dilaton theory and static black hole}
We examine the Einstein-Maxwell-Yang-Mills-dilaton theory under assumption that the dilaton, Maxwell and Yang-Mills fields are minimally coupled to gravity, whereas the couplings between dilaton and Yang-Mills or Maxwell fields are nonminimal. The action for this theory in $n+1$--dimensional space-time takes the form:
\begin{eqnarray}\label{action}
 S=\frac{1}{16\pi}\int_{{\cal M}} {\rm d}^{n+1}x\sqrt{-g}\left(R-\frac{4}{n-1}\nabla^{\mu}\Phi\nabla_{\mu}\Phi-V(\Phi)-e^{-\frac{4\alpha\Phi}{n-1}}Tr(F^{(a)}_{\mu\nu}F^{(a)\mu\nu})-e^{-\frac{4\be\Phi}{n-1}}\cal{F}_{\mu\nu}\cal{F}^{\mu\nu}\right)+S_{GHY},
\end{eqnarray}
where $g$ is the determinant of the metric tensor $g_{\mu\nu}$, $R$ is the scalar curvature, $\Phi$ is the dilaton field, $V(\Phi)$ denotes the dilaton potential, $F^{(a)}_{\mu\nu}$ and $\cal{F}_{\mu\nu}$ denote the Yang-Mills and Maxwell fields respectively, the parameters $\al$ and $\be$ are the dilaton-Yang-Mills and dilaton-Maxwell coupling constants correspondingly. The second term in the action (\ref{action}), namely $S_{GHY}=\frac{1}{8\pi}\int_{\partial{\cal M}} d^{n}x\sqrt{-h}K$ is the boundary Gibbons-Hawking-York (GHY) term, introduced to have the variational problem well-posed. In the GHY-term $h$ is the determinant of the boundary metric $h_{\mu\nu}$ and $K$ is the trace of the extrinsic curvature tensor. Here we point out that in the action (\ref{action}) the gravitational constant $G$ is set to one ($G=1$).

The Yang-Mills gauge field is defined in the standard way, namely:
\begin{equation}\label{YM_field}
F^{(a)}_{\mu\nu}=\partial_{\mu}A^{(a)}_{\nu}-\partial_{\nu}A^{(a)}_{\mu}+\frac{1}{2\bar{\sigma}}C^{(a)}_{(b)(c)}A^{(b)}_{\mu}A^{(c)}_{\nu},
\end{equation} 
where $A^{(a)}_{\mu\nu}$ are the components of the Yang-Mills potential, $C^{(a)}_{(b)(c)}$ are the structure constants for the corresponding gauge group and $\bar{\sigma}$ is the coupling constant for the field. Here we suppose that the gauge group is $SO(n)$.

The Maxwell field is also defined in the standard way:
\begin{equation}\label{Max_field}
{\cal{F}}_{\mu\nu}=\partial_{\mu}{\cal{A}}_{\nu}-\partial_{\nu}{\cal{A}}_{\mu},
\end{equation}
and here ${\cal{A}}_{\mu}$ are the components of the electromagnetic field potential.

Varying the action (\ref{action}) with respect to the metric $g_{\mu\nu}$, dilaton field $\Phi$, Yang-Mills gauge potential $A^{(a)}_{\mu\nu}$ and electromagnetic field potential $A_{\mu}$ we arrive at the equations of motion for the system:
\begin{eqnarray}\label{einstein}
\nonumber R_{\mu\nu}=\frac{g_{\mu\nu}}{n-1}\left(V(\Phi)-e^{-4\al\Phi/(n-1)}Tr(F^{(a)}_{\rho\sigma}F^{(a)\rho\sigma})-e^{-4\be\Phi/(n-1)}\cal{F}_{\rho\sigma}\cal{F}^{\rho\sigma}\right)\\
+\frac{4}{n-1}\partial_{\mu}\Phi\partial_{\nu}\Phi+2e^{-4\al\Phi/(n-1)}
Tr(F^{(a)}_{\mu\sigma}{F^{(a)\sigma}_{\nu}})+2e^{-4\be\Phi/(n-1)}\cal{F}_{\mu\sigma}{\cal{F}_{\nu}}^{\sigma};
\end{eqnarray}
\begin{equation}\label{scal_eq}
\nabla_{\mu}\nabla^{\mu}\Phi=\frac{n-1}{8}\frac{\partial V}{\partial \Phi}-\frac{\al}{2}e^{-4\al\Phi/(n-1)}Tr(F^{(a)}_{\rho\sigma}F^{(a)\rho\sigma})-\frac{\be}{2}e^{-4\be\Phi/(n-1)}\cal{F}_{\mu\nu}\cal{F}^{\mu\nu};
\end{equation}
\begin{equation}\label{YM_eq}
\nabla_{\mu}(e^{-4\al\Phi/(n-1)}F^{(a)\mu\nu})+\frac{1}{\bar{\sigma}}e^{-4\al\Phi/(n-1)}C^{(a)}_{(b)(c)}A^{(b)}_{\mu}F^{(c)\mu\nu}=0;
\end{equation}
\begin{equation}\label{em_eq}
\nabla_{\mu}(e^{-4\be\Phi/(n-1)}{\cal{F}}^{\mu\nu})=0.
\end{equation}
In this work we obtain and examine a static black hole's solution, thus the metric can be written in the form:
\begin{equation}\label{metric}
 ds^2=-W(r)dt^2+\frac{dr^2}{W(r)}+r^2R^2(r)d\Omega^2_{n-1},
\end{equation}
where $d\Omega^2_{n-1}$ is the line element of $n-1$-dimensional unit hypersphere.

For the Yang-Mills potential $A^{(a)}_{\mu}$ we utilize Wu-Yang ansatz and take the components of the gauge potential in the following form:
\begin{equation}\label{gauge_pot}
{\bf A}^{(a)}=\frac{\bq}{r^2}C^{(a)}_{(i)(j)}x^{i}dx^{j}, \quad r^2=\sum^{n}_{j=1}x^2_j,
\end{equation}
where indices $a$, $i$ and $j$ run the following ranges $2\leqslant j+1<i\leqslant n$ and $1\leqslant a\leqslant n(n-1)/2$. We note here that the parameter $\bar{q}$ in the written above gauge potential is taken to be equal to the coupling constant $\bar{\sigma}$, namely $\bar{q}=\bar{\sigma}$.  

Here we would like to point out that the Wu-Yang ansatz (\ref{gauge_pot}) written above is a bit restrictive and it allows to derive just some subclass of spherically symmetric solutions of this theory. Even in those pioneering \cite{Bartnik_PRL88,Bizon_PRL90,Volkov_JETP89,Kuenzle_JMP90} as well as in later \cite{Brihaye_PRD07} papers a more general ansatz for the Yang-Mills gauge potential was used, which included additional radial function and the latter one might be derived as a solution of the equations of motion (\ref{einstein})-(\ref{em_eq}) together with the metric function $W(r)$ and the potential for electromagnetic field, but as far as we know in this more general case only numerical solutions were obtained \cite{Brihaye_PRD07}. In this work we are going to derive an exact analytical solution and Yang-Mills ansatz (\ref{gauge_pot}) allows us to achieve this aim. We would like to emphasize here that the Wu-Yang ansatz (\ref{gauge_pot}) was utilized to derive mainly static black hole solutions which were mentioned in the previous section, although recently by virtue of this ansatz and Newmann-Janis algorithm a rotating black hole solution  was obtained \cite{Ghosh_EPJC14}. 
 
The coordinates $x_i$ in the relation (\ref{gauge_pot}) can be defined by virtue of the angular variables, namely for them we have the following relations:
\begin{eqnarray}
\nonumber x_1=r\cos{\chi_{n-1}}\sin{\chi_{n-2}}\ldots\sin{\chi_1},\quad x_2=r\sin{\chi_{n-1}}\sin{\chi_{n-2}}\ldots\sin{\chi_1},\\
\nonumber  x_3=r\cos{\chi_{n-2}}\sin{\chi_{n-3}}\ldots\sin{\chi_1},\quad x_4=r\sin{\chi_{n-2}}\sin{\chi_{n-3}}\ldots\cos{\chi_1},\\
\nonumber \cdots \quad\quad\\
x_n=r\cos{\chi_1}
\end{eqnarray}
Having used the angular variables $\chi_1,\ldots,\chi_{n-1}$  we can represent the line element of the unit hypersphere of the metric (\ref{metric}) in the form:
\begin{equation}
d\Omega^2_{n-1}=d\chi^2_{1}+\sum^{n-1}_{j=2}\prod^{j-1}_{i=1}\sin^2{\chi_{i}}d\chi^2_{j},
\end{equation}
Since we are going to obtain a static solution, we can choose the electromagnetic potential form as follows: ${\cal{A}}={\cal{A}}(r)dt$. Thus, taking the written above form of the electromagnetic potential and the metric (\ref{metric}) into account we can find the electromagnetic field tensor, which takes the form:
\begin{equation}\label{EM_field}
{\cal{F}}_{tr}=\frac{q}{(rR)^{n-1}}e^{4\be\Phi/(n-1)},
\end{equation}
where $q$ is an integration constant related to the electric charge of the black hole. Using the obtained relation for the electromagnetic field we calculate the Maxwell field invariant which is given in the equations of motion (\ref{einstein})-(\ref{scal_eq}):
\begin{equation}
{\cal{F}}_{\mu\nu}{\cal{F}}^{\mu\nu}=-2\frac{q^2}{(rR)^{2(n-1)}}e^{8\be\Phi/(n-1)}
\end{equation} 
Using the Wu-Yang ansatz for the nonabelian potential (\ref{gauge_pot}) it is easy to verify that the corresponding field equations (\ref{YM_eq}) are  satisfied. Here we also write the invariant for the Yang-Mills field:
\begin{equation}
Tr(F^{(a)}_{\rho\sigma}F^{(a)\rho\sigma})=(n-1)(n-2)\frac{q^2}{r^4R^4}.
\end{equation}
To derive a solution of the equations of motion for the gravitational (\ref{einstein}) and the dilaton (\ref{scal_eq}) fields one should define an evident form for the dilaton potential $V(\Phi)$. Here we choose this potential in the so-called Liouville form:
\begin{equation}\label{dil_pot}
V(\Phi)=\Lambda e^{\lambda\Phi}+\Lambda_1 e^{\lambda_1\Phi}+\Lambda_2e^{\lambda_2\Phi}+\Lambda_3e^{\lambda_3\Phi},
\end{equation} 
and the parameters $\l$, $\l_i$ and $\L_i$ ($i=1,2,3$) can be derived from the equations of motion (\ref{einstein})-(\ref{scal_eq}). We point out here that the Liouville-type potentials were used in numerous papers where black holes with dilaton fields were studied \cite{Sheykhi_PRD07,Radu_CQG05,Stetsko_EYM20,Stetsko_EPJC19}. We also make use of a specific ansatz for the function $R(r)$ which was extensively utilized in the works where dilaton fields with Liouville potential were considered:
\begin{equation}\label{ansatz_R}
R(r)=e^{2\al\Phi/(n-1)}.
\end{equation} 
Taking into account the given above ansatz (\ref{ansatz_R}) and the chosen form for the dilaton potential (\ref{dil_pot}) we solve the field equations (\ref{einstein})-(\ref{scal_eq}) and write the metric function $W(r)$ in the form:
\begin{eqnarray}
\nonumber W(r)=-mr^{1+(1-n)(1-\gamma)}+\frac{(n-2)(1+\al^2)^2}{(1-\al^2)(\al^2+n-2)}b^{-2\gamma}r^{2\gamma}-\frac{\Lambda(1+\al^2)^2}{(n-1)(n-\al^2)}b^{2\gamma}r^{2(1-\gamma)}+\\\frac{(n-2)(1+\al^2)^2\bar{q}^2}{(\al^2-1)(n+3\al^2-4)}b^{-6\gamma}r^{6\gamma-2}+\frac{2(1+\al^2)^2q^2}{(2-n+\al^2-2\al\be)(1-n+\al^2-\al\be)}b^{(\k-2(n-1))\g}r^{2-2(n-1)(1-\g)-\k\g}\label{metric_W},
\end{eqnarray}
where $\g=\al^2/(1+\al^2)$ and $\k=2\be/\al$. We point out here that $m$ is an integration constant which, as it will be shown below, is related to the mass of the black hole and $b$ is also an integration constant which is supposedly related to some rescaling properties of the obtained solution, but it seems to be no physical reason to fix it in any way. Apart of the metric function (\ref{metric_W}) the field equations (\ref{einstein})-(\ref{scal_eq}) also give rise to the dilaton field $\Phi$ of the form:
\begin{equation}
\Phi(r)=\frac{\al(n-1)}{2(1+\al^2)}\ln{\left(\frac{b}{r}\right)}.
\end{equation}
It is worth noting that the given above relation for the dilaton field is not the most general solution of the field equations, but as it was pointed out in the earlier works this form allows to derive quite general conclusions. It has been noted above that the parameters of the Liouville potential (\ref{dil_pot}) can be chosen to fulfill the equations of motion (\ref{einstein})-(\ref{scal_eq}). Here we take them in the form:
\begin{eqnarray}
\lambda=\frac{4\al}{(n-1)},\quad \lambda_1=\frac{4}{\al(n-1)}, \quad \lambda_2=\frac{4(2-\al^2)}{\al(n-1)},\quad \l_3=\frac{4(n-1+\al\be)}{\al(n-1)}, \label{param_1}\\
\Lambda_1=\frac{\alpha^2(n-1)(n-2)}{\al^2-1}b^{-2}, \quad \Lambda_2=-\frac{\al^2(n-1)(n-2)}{\al^2-1}\bar{q}^2b^{-4},\quad \L_3=\frac{2\al(\al-\be)q^2}{n-1+\al\be-\al^2}b^{2(1-n)}  \label{param_2}.
\end{eqnarray}     
The parameter $\L$ in the dilaton potential $V(\Phi)$ and consequently in the metric can take arbitrary value and it is treated as an effective cosmological constant. We point out here that similar situation happens for Einstein-Maxwell-dilaton \cite{Stetsko_EPJC19} and Einstein-Yang-Mills-dilaton black holes \cite{Stetsko_EYM20}. We also note that the metric (\ref{metric_W}) is singular at the points $\al=1$ (the so-called string singularity) and if $n=\al^2$, here we also have completely identical situation to the mentioned Einstein-Maxwell-dilaton and Einstein-Yang-Mills-dilaton black holes.  We note that in the following we assume that $\al<1$, the opposite case will be considered elsewhere.

\begin{figure}
\centerline{\includegraphics[scale=0.3,clip]{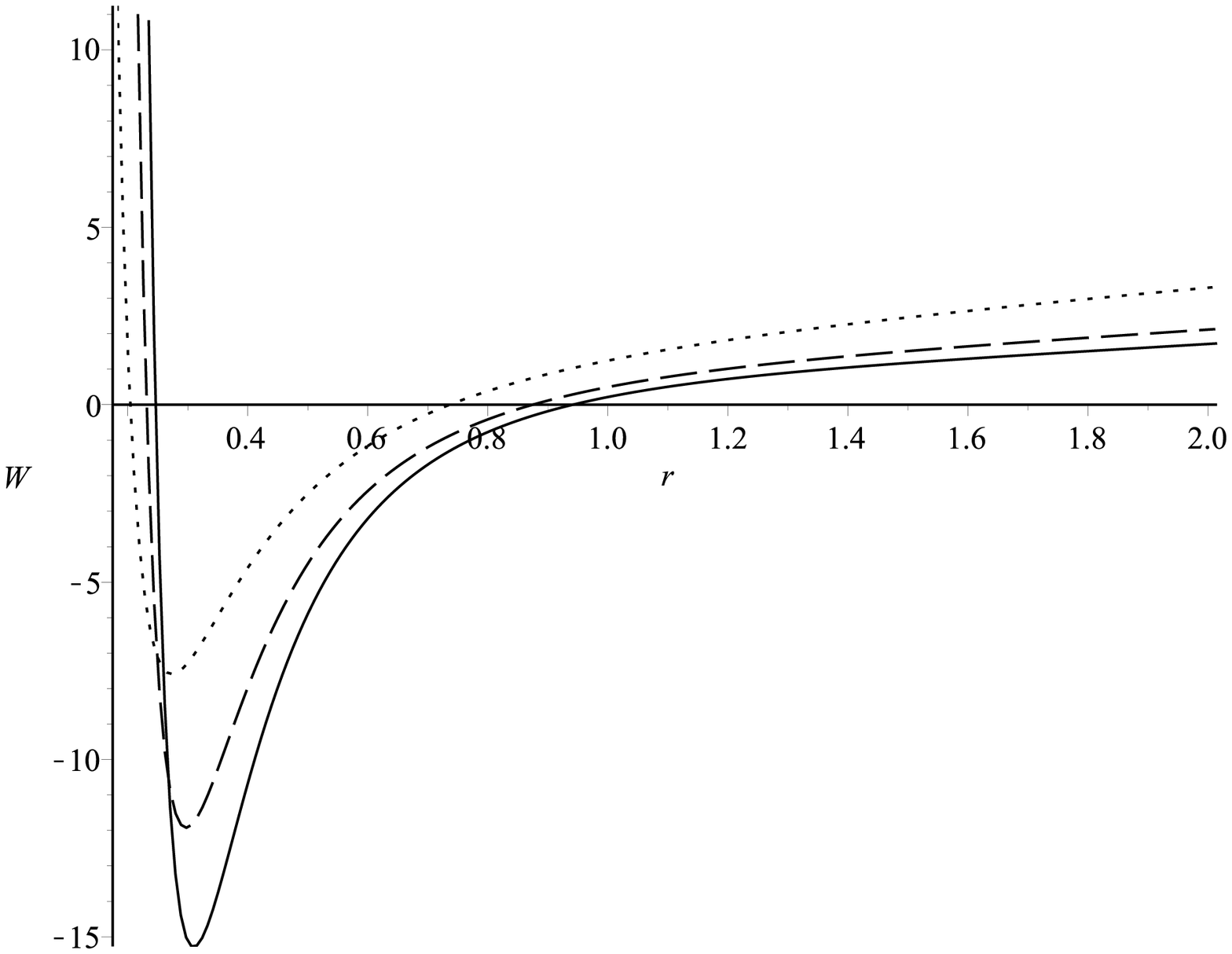}\includegraphics[scale=0.3,clip]{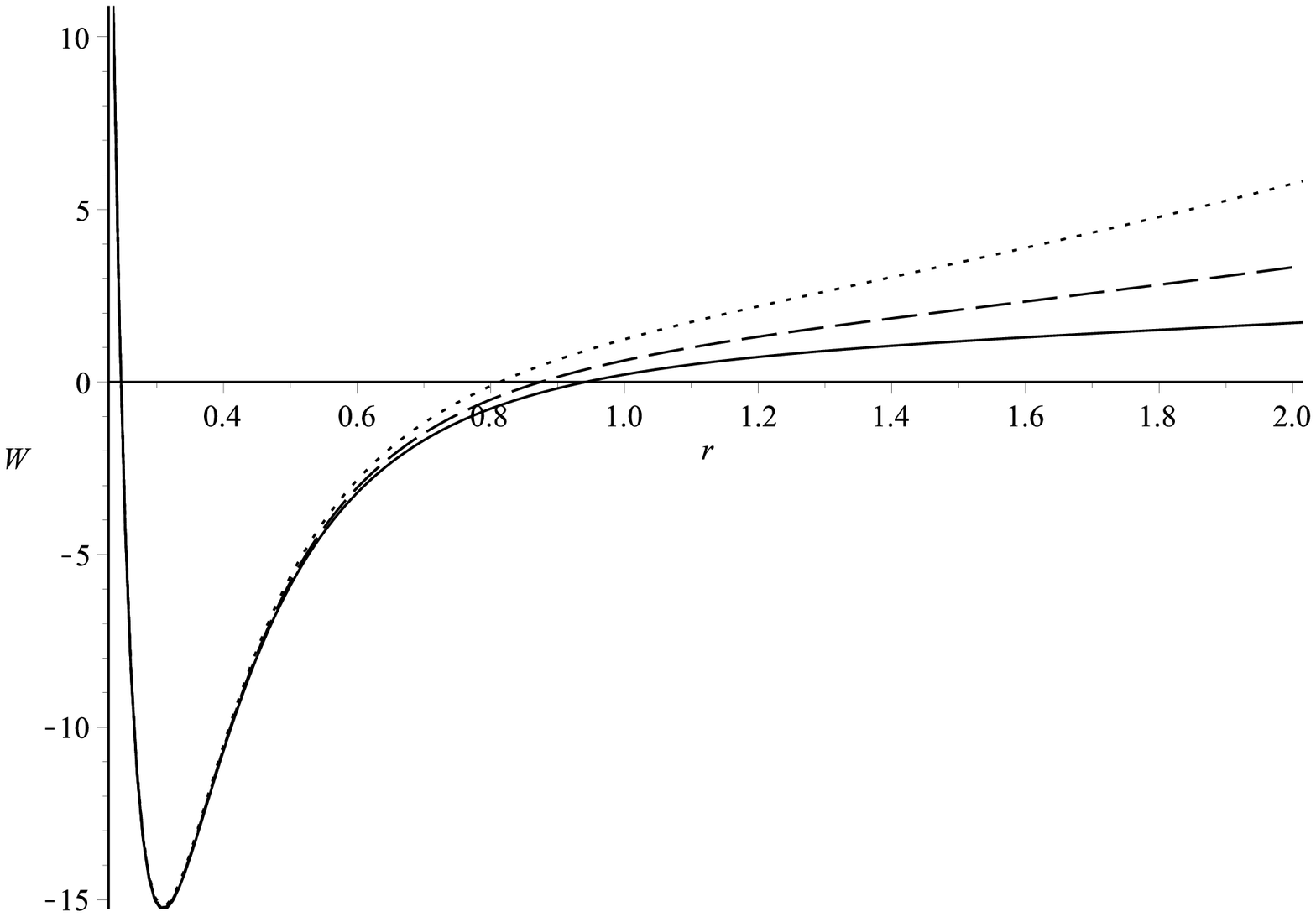}}
\caption{Metric function W(r) for different values of parameter $\al$ (the left graph) and cosmological constant $\Lambda$ (the right one). For both graphs we have taken: $n=5,\beta=0.1,b=1,m=1,\bar{q}=0.2,q=0.3$. For the left plot the parameter $\al$ is chosen to be: $\al=0.1$, $\al=0.3$, $\al=0.5$ for solid, dashed and dotted curves respectively and $\L=-4$. For the right graph the varied parameter $\L$ is taken to be $\L=-4$, $\L=-12$ and $\L=-24$ for solid, dashed and dotted lines correspondingly, while $\al=0.1$.}\label{metric_graph}
\end{figure}

The evident form of the metric function $W(r)$ (\ref{metric_W}) is quite complicated, but nevertheless we would like to remark some important features of this function. Firstly, we would like to point out that at the infinity the dominant term is proportional to $\sim r^{2(1-\gamma)}$, thus the metric of the black hole is not asymptotically flat. In addition, if $\L$ is negative there is no cosmological horizon similarly as in AdS case, but here the metric function is not exactly anti-de Sitterian, AdS-like behaviour of the metric can be recovered just in case when $\al\rightarrow 0$. For very small distances (close to the origin) the behaviour of the metric function is mainly driven by the term $\sim q^2 r^{2-2(n-1)(1-\g)-\k\g}$, the term becomes infinite when $r\rightarrow 0$, so this term mainly defines the character of singularity at the origin, but to be exact, the other terms, for instance the so-called Schwarzschild term $\sim mr^{1+(1-n)(1-\gamma)}$, are also singular at the origin, but its singularity is weaker and it cannot change the character of the singularity at the origin. To comprehend the behaviour of the metric function deeply we give graphical dependence $W(r)$.  Namely, the Figure [\ref{metric_graph}] shows the considerable influence of the parameter $\al$ for all distances, whereas the variation of the parameter $\L$ is substantial mainly for large distances, since the term containing $\L$ defines the behaviour of the metric at large distances. We also point out that the influence of the variation of the parameter $\be$ becomes visible for small distances, since this parameter is present only in the term which is dominant for small distances. The Figure [\ref{metric_graph}] shows that the black hole might have two horizons, and the outer of them is the event horizon, but it should be also pointed out that the increase of the parameter $q$, when the other parameters such as $m$ and $\bar{q}$ are held fixed, gives rise to the situation when the horizons points become closer and closer, the following increase of the parameter $q$ leads to the merging of these points in the so called degenerate case (or extreme case) and finally, further increase of the charge $q$ leads to the appearance of a naked singularity.  

Since the metric (\ref{metric}) is singular at several points, namely at the infinity, at the horizons and at the origin, it is important to find out what is the character of singularity points. To perform this task we examine the Kretschmann scalar which in general has the following form:
\begin{equation}\label{Kr_scal}
R_{\mu\nu\rho\sigma}R^{\mu\nu\rho\sigma}=\left(W''\right)^2+\frac{(n-1)}{(rR)^2}\left[\left((rR)'\right)^2\left(W'\right)^2+\left(2(rR)''W-(rR)'W'\right)^2\right]+\frac{2(n-1)(n-2)}{(rR)^4}\left(1-\left((rR)'\right)^2W\right)^2.
\end{equation} 
Since the form of the metric function (\ref{metric_W}) is not very simple we analyze the behaviour of the Kretschmann scalar just in the vicinity of the singularity points. Firstly, it can be shown that at the horizons the Kretschmann scalar (\ref{Kr_scal}) is not singular, so these points are just the points of coordinate singularity as it should be for a black hole's solution. To investigate the behaviour of the metric at the origin we just take the leading term of the metric (\ref{metric_W}) and as a result we can write:
\begin{equation}\label{kr_sc_inf}
R_{\mu\nu\rho\sigma}R^{\mu\nu\rho\sigma}\simeq f(n,\al,\be)q^4r^{-4(n-1)(1-\gamma)-2\k\g},
\end{equation}
and here $f(n,\al,\be)$ is some function of dimension $n$ and coupling parameters $\al$ and $\be$, the evident form of this function  can be written, but we do not give it here since the explicit expression for this function is not important here. The evident form of the Kretschmann scalar in the vicinity of the origin  shows that this point is the only point of true physical singularity.

At the infinity the leading term of the metric function (\ref{metric_W}) is the same as for the Einstein-Maxwell-dilaton \cite{Stetsko_EPJC19} or Einstein-Yang-Mills-dilaton \cite{Stetsko_EYM20} cases, thus the Kretschmann scalar also has the same behaviour:
\begin{equation}
R_{\mu\nu\rho\sigma}R^{\mu\nu\rho\sigma}\simeq\frac{2\L^2}{(n-1)^2(n-\al^2)^2}\left(2(1-\al^2)^2+n(n-1)+2(n-1)(1+\al^2)^2\right)b^{4\gamma}r^{-4\gamma}.
\end{equation}
It is easy to conclude that at the infinity the Kretschmann scalar goes to zero, but in the limit $\al=0$ we recover the form obtained for AdS case.
\section{Thermodynamics of the black hole}   
\begin{figure}
\centerline{\includegraphics[scale=0.3,clip]{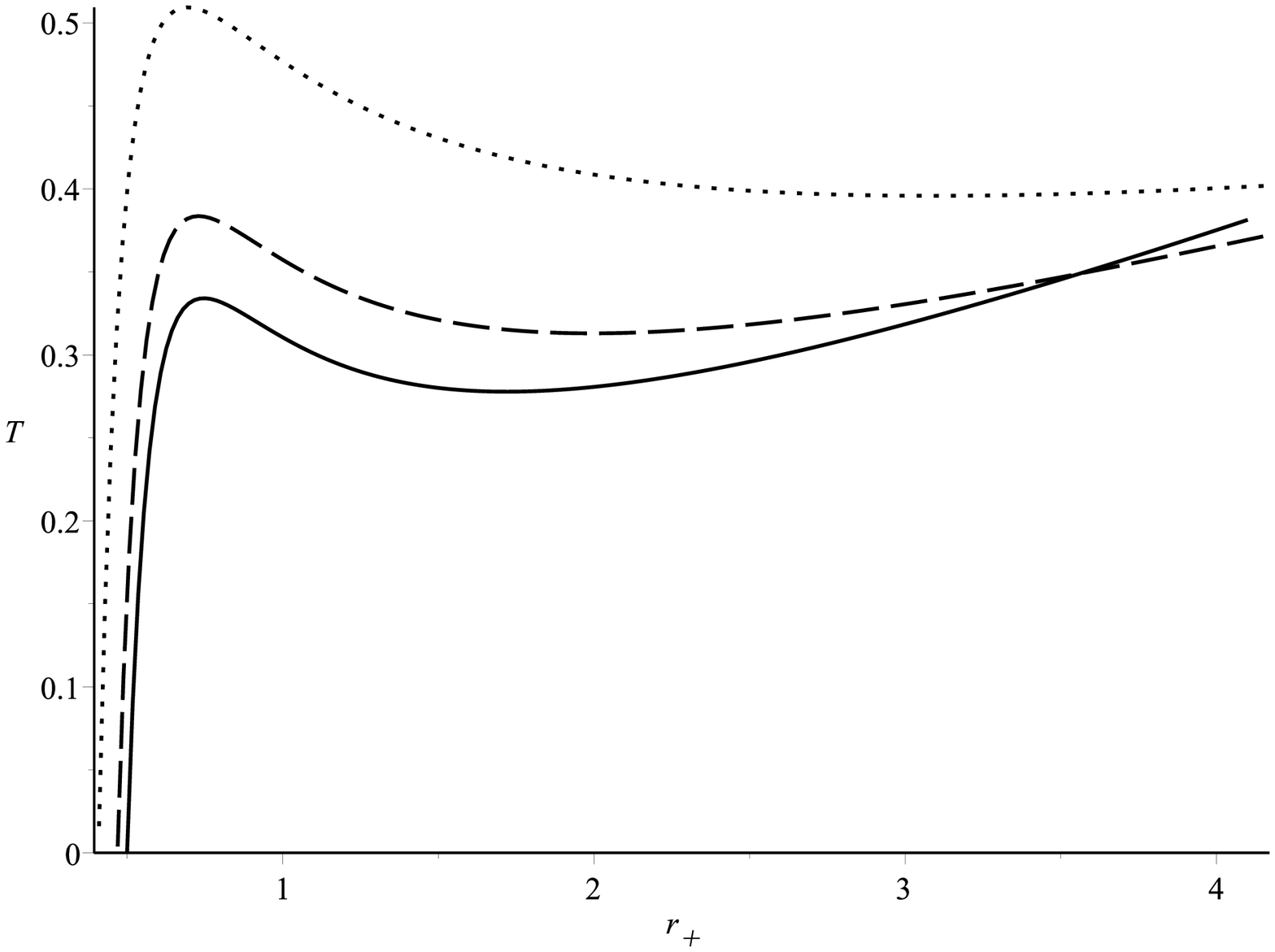}\includegraphics[scale=0.3,clip]{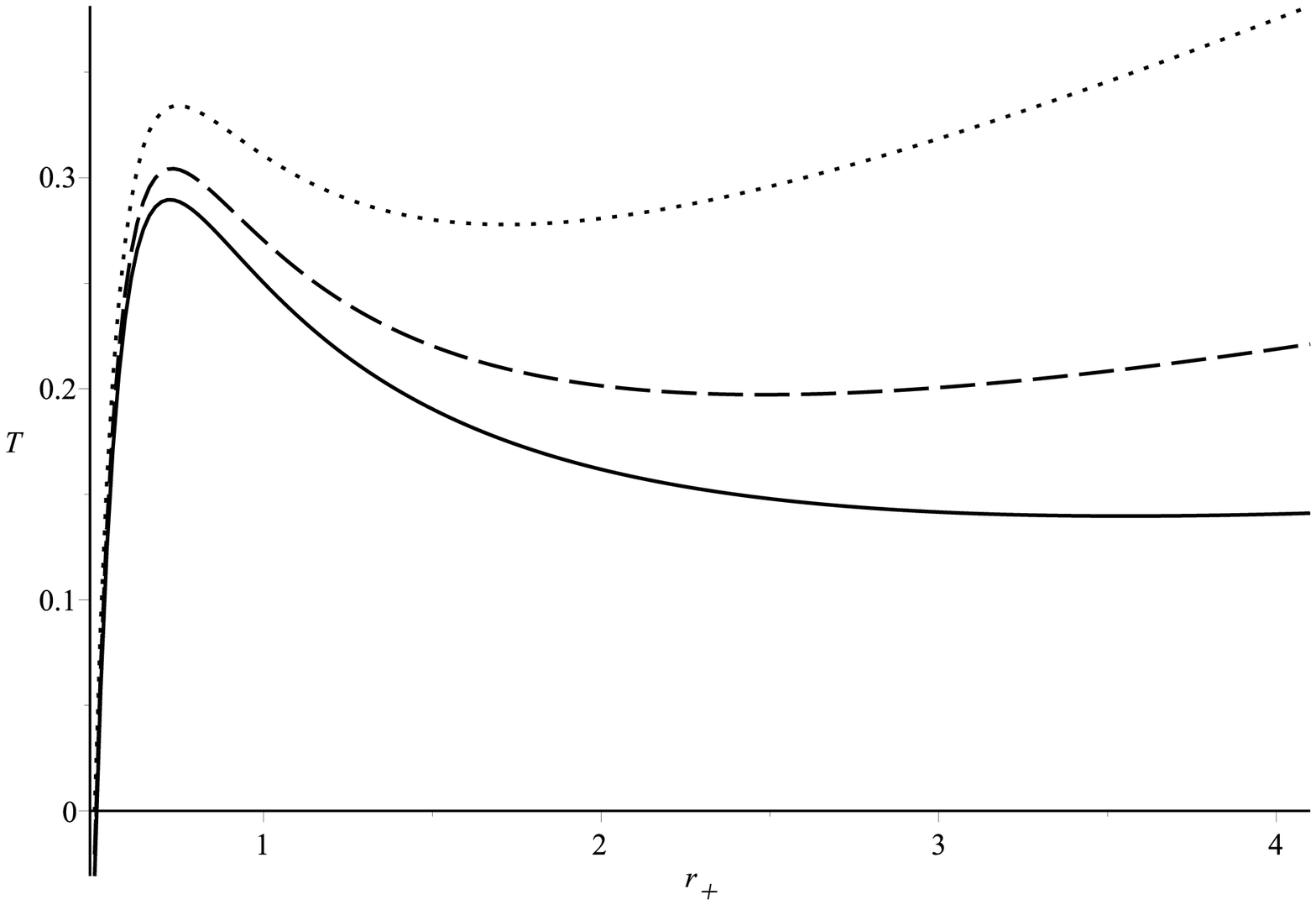}}
\caption{Temperature $T$ as a function of the horizon radius $r_+$ for various values of the parameter $\al$ (the left graph) and cosmological constant $\L$ (the right one). For both graphs we have taken: $n=5,\beta=0.15,b=1,\bar{q}=0.2,q=0.3$. For the left plot the parameter $\al$ is chosen to be: $\al=0.1$, $\al=0.3$, $\al=0.5$ for solid, dashed and dotted curves respectively and $\L=-4$. For the right graph the varied parameter $\L$ is taken to be $\L=-1$, $\L=-2$ and $\L=-4$ for solid, dashed and dotted lines correspondingly, while $\al=0.1$.}\label{temp_graph}
\end{figure}

We start this section calculating the temperature of the black hole. It is known that the definition of the black hole's temperature is based on the concept of surface gravity which is defined as follows:
\begin{equation}
\varkappa^2=-\frac{1}{2}\nabla_a\chi_b\nabla^a\chi^b, 
\end{equation}
where $\chi^a$ represents a Killing vector field which should be null at the horizon. For a static black hole's solution the Killing vector can be chosen to be the time translation vector $\chi^{a}=\frac{\partial}{\partial t}$. As a result, the temperature can be written in the form:
\begin{eqnarray}\label{temp}
\nonumber T=\frac{\varkappa}{2\pi}=\frac{(1+\al^2)}{4\pi}\left(\frac{n-2}{1-\al^2}b^{-2\gamma}r^{2\gamma-1}_{+}\left(1-\bar{q}^2b^{-4\gamma}r^{2(2\gamma-1)}_{+}\right)\right.\\\left.-\frac{\Lambda}{n-1}b^{2\gamma}r_{+}^{1-2\gamma}+\frac{2q^2}{1-n+\al^2-\be\al}b^{(\k-2(n-1))\gamma}r^{1-2(n-1)(1-\gamma)-\k\gamma}_{+}\right),
\end{eqnarray}
and here $r_+$ is the radius of the event horizon of the black hole. The dependence $T=T(r_+)$ (\ref{temp}) shows that for large radius of the horizon the leading term in the function is $\sim r^{1-2\g}_+$ and in the limit $\al=0$ for large radii of the horizon we have de-Sitterian or anti-de Sitterian linear dependence. For some intermediate values of the radius of the horizon  the function might be nonmonotonous since the contribution of different terms might be of the same order. Finally, for small radius of the horizon the dominant term in the temperature is the term caused by the electromagnetic part (the fourth term in (\ref{temp})) and this is natural since for the metric function $W(r)$ (\ref{metric_W}) for small distances we also had the dominance of the term  caused by the Maxwell field. Figure [\ref{temp_graph}] demonstrates graphical dependence of the function $T=T(r_+)$. From this figure we can conclude that the variation of the parameter $\al$ has its impact on the temperature in all the range of the variation of the horizon radius, whereas the variation of the cosmological constant $\L$ leads to considerable change of the temperature just for large radius of the horizon. It should be noted that the variation of the parameter $\be$ becomes substantial for small radii of the horizon, since the corresponding term of the metric function is also substantial for small distances. It is worth pointing out that nonmonotonous character of the temperature means some kind of critical behaviour for the black hole and some aspects of this criticality will be examined below. 

To develop thermodynamics of the black hole we have to define its entropy. The entropy can be derived in several ways, in particular by virtue of well-established Wald's approach \cite{Wald_PRD93}. The Wald entropy is defined as follows:
\begin{equation}\label{wald_S}
S=-2\pi\int_{\Sigma}d^{n-1}x\sqrt{h}\frac{\partial{\cal L}}{\partial R_{\mu\nu\k\l}}\hat{\ve}_{\mu\nu}\hat{\ve}_{\k\l},
\end{equation}
where $\Sigma$ denotes a Killing horizon, $h$ is the determinant of the induced metric on the horizon surface, ${\cal L}$ is the gravitational Lagrangian and $\hat{\ve}_{\mu\nu}$ is the binormal to the horizon, normalized by the condition $\hat{\ve}_{\mu\nu}\hat{\ve}^{\mu\nu}=-2$. Taking the derivative of the Lagrangian ${\cal L}$ we obtain:
\begin{equation}
\frac{\partial{\cal L}}{\partial R_{\mu\nu\k\l}}=\frac{1}{32\pi}\left(g^{\mu\k}g^{\nu\l}-g^{\mu\l}g^{\nu\k}\right).
\end{equation}
Substituting the given above derivative into the relation (\ref{wald_S}) we can calculate the entropy of the black hole which  can be written in the form:
\begin{equation}\label{entropy}
S=\frac{\omega_{n-1}}{4}b^{(n-1)\g}r^{(n-1)(1-\gamma)}_+,
\end{equation}
where $\omega_{n-1}$ is the surface area of a $n-1$--dimensional unit hypersphere. We point out here that the relation for the entropy (\ref{entropy}) (so-called area law) is expectable, since the gravitational part of the Lagarangian is given by the Ricci scalar and the coupling between gravity and other fields is minimal.


Since our black hole has an electric charge which belongs to the so-called global charges it means that the electric charge might be a thermodynamic value similarly as it takes place for the Einstein-Maxwell-dilaton or even ordinary RN black holes. The electric charge of the black holes can be obtained by virtue of the relation:
\begin{equation}
Q=\frac{1}{4\pi}\int \exp\{-4\be\Phi/(n-1)\}*F
\end{equation} 
and here $*F$ denotes the form dual to the electromagnetic field form. We also note that the integration in the above relation should be performed over a closed spacelike hypersurface which encloses the black hole. Having calculated the above integral we obtain:
\begin{equation}\label{el_charge}
Q=\frac{\omega_{n-1}}{4\pi}q.
\end{equation} 
The other parameters which are given in the metric function $W(r)$ are held fixed and here we suppose that the first law of black hole's thermodynamics takes the form:
\begin{equation}\label{first_law}
dM=TdS+\Phi_e dQ,
\end{equation}
where:
\begin{equation}
T=\left(\frac{\partial M}{\partial S}\right)_Q, \quad \Phi_e=\left(\frac{\partial M}{\partial Q}\right)_S,
\end{equation}
and here $\Phi_e$ denotes the electric potential which is conjugate to the charge. In the first law (\ref{first_law}) $dM$ is the variation of the mass of the black hole. The first law allows us to obtain the thermodynamic mass of the black hole, which can be written in the form:
\begin{equation}\label{mass_bh}
M=\frac{(n-1)b^{(n-1)\gamma}\omega_{n-1}}{16\pi(1+\al^2)}m.
\end{equation} 
We point out here that the mass we have just obtained is completely defined by the integration constant $m$, the so-called mass parameter. Similar expression for the mass we derived for Einstein-Maxwell-dilaton \cite{Stetsko_EPJC19} and Einstein-Yang-Mills-dilaton \cite{Stetsko_EYM20} black holes.

\begin{figure}
\centerline{\includegraphics[scale=0.3,clip]{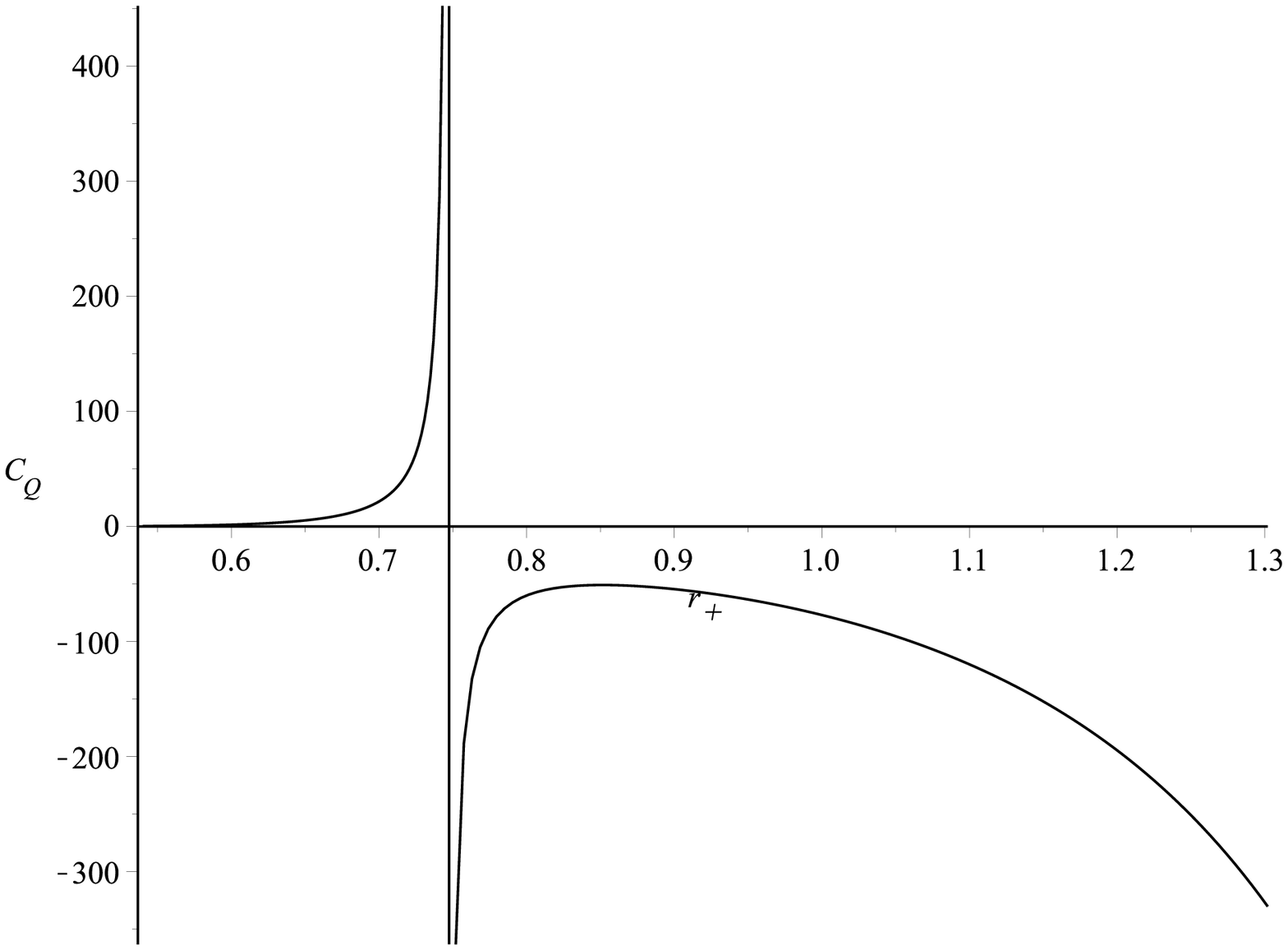}\includegraphics[scale=0.3,clip]{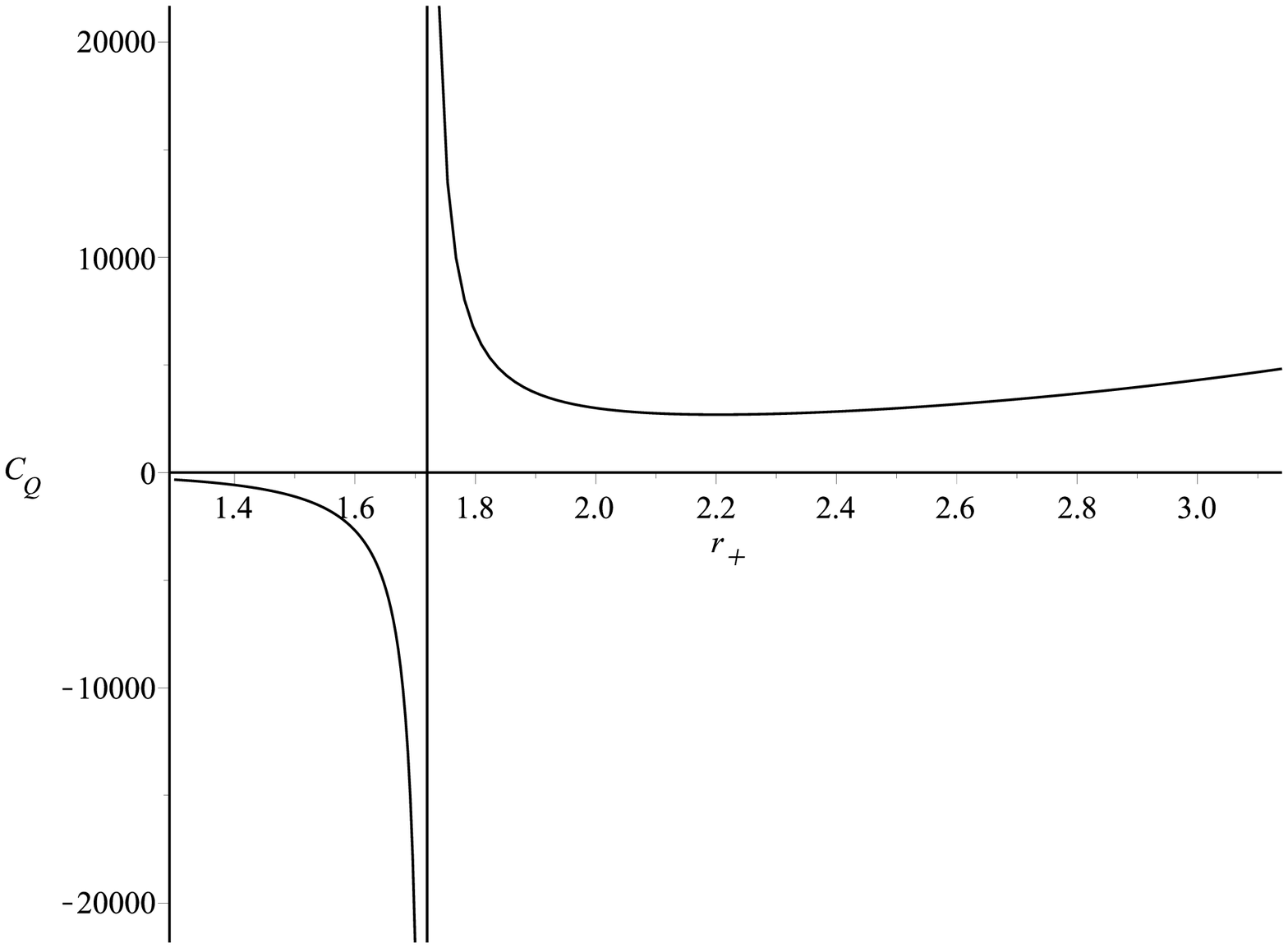}}
\caption{Heat capacity $C_{Q}$ as a function of the horizon radius $r_+$ which shows two discontinuity points. The fixed parameters are chosen to be: $n=5,\al=0.1,\be=0.15,\L=-4,b=1,\bar{q}=0.2,q=0.3$.  }\label{Heat_cap_gr}
\end{figure}

One of the most important tasks in black hole thermodynamics is the investigation of thermal stability of the black hole's solution. To examine the thermal stability we calculate heat capacity which is defined as follows:
\begin{equation}\label{heat_capacity}
C_{Q}=T\left(\frac{\partial S}{\partial T}\right)_{Q}=T\left(\frac{\partial S}{\partial r_+}\right)_{Q}\left(\frac{\partial T}{\partial r_+}\right)^{-1}_{Q}.
\end{equation} 
Having performed some elementary calculations we can write:
\begin{eqnarray}
\nonumber C_Q=\frac{(n-1)\omega_{n-1}}{4}b^{(n-1)\g}r^{(n-1)(1-\gamma)}_+\left(\frac{n-2}{1-\al^2}b^{-2\gamma}r^{2\gamma-1}_+\left(1-\bar{q}^2b^{-4\gamma}r^{2(2\gamma-1)}_{+}\right)-\right.\\\nonumber\left.\frac{\Lambda}{n-1}b^{2\gamma}r_+^{1-2\gamma}+\frac{2q^2b^{(\k-2(n-1))\gamma}}{1-n+\al^2-\be\al}r_+^{1-2(n-1)(1-\gamma)-\k\gamma}\right)\left[(2-n)b^{-2\gamma}r^{2(\gamma-1)}_{+}\left(1-3\bar{q}^2b^{-4\gamma}\times\right.\right.\\\left.\left.r^{2(2\gamma-1)}_{+}\right)-\frac{\Lambda(1-\alpha^2)}{n-1}b^{2\gamma}r_{+}^{-2\gamma}+\frac{2q^2(3-2n+\al^2-2\be\al)}{1-n+\al^2-\be\al}b^{(\k-2(n-1))\gamma}r_{+}^{-2(n-1)(1-\gamma)-\k\gamma}\right]^{-1}.\label{heat_capac}
\end{eqnarray} 
As it was noted above, the nonmonotonous behaviour of the temperature leads to the conclusion about discontinuity of the heat capacity. The discontinuity of the heat capacity reflects the appearance of Hawking-Page phase transition and the points where it takes place separate stable and unstable domains. The Figure [\ref{Heat_cap_gr}] shows these discontinuities. It is easy to conclude that for relatively large radii of the horizon the black hole is stable as a thermodynamic system, then for some intermediate values of $r_+$, namely between the discontinuity points there is the domain of instability and below the lower point of discontinuity we again have the  stable domain. With the increase of the absolute value of the cosmological constant $\L$ the dependence $T=T(r_+)$ turns to be less nonmonotonous and finally the nonmonotonicity disappears. At the same time analyzing the behaviour of the heat capacity $C_{Q}$ with the increase of the module of $\L$ we can conclude that the discontinuity points are getting closer and closer and in the end they merge to one point with the following transformation of the discontinuity into a peak, which also loses its height with the increase of the module of $\L$. We also point out here that from the qualitative point of view the behaviour of the heat capacity $C_Q$ is completely identical to the situations which take place for Einstein-Maxwell-dilaton \cite{Stetsko_EPJC19} and Einstein-Yang-Mills-dilaton \cite{Stetsko_EYM20} cases.

\section{Thermodynamics in the extended phase space}
The concept of the extended thermodynamic phase space has gained considerable interest for recent years \cite{Kubiznak_JHEP12,Kubiznak_CQG17}. The main idea of this approach is based on the assumption about the variation of the cosmological constant, namely the cosmological constant is supposed to be related to the thermodynamic pressure\cite{Kastor_CQG09,Cvetic_PRD11,Dolan_CQG11}. It should be also pointed out that in various cosmological models, where stress-energy of a fluid is used, the cosmological constant might be treated almost on the same footing as the thermodynamic pressure which is present in the stress-energy tensor of the fluid. The concept of extended phase space allows to develop deep analogy between thermodynamics and critical phenomena of ordinary condensed matter systems from one side and black holes' thermodynamics from the other one \cite{Kubiznak_CQG17}.  

We introduce the thermodynamic pressure similarly as it is done for the Einstein-Maxwell-dilaton \cite{Stetsko_EPJC19} and Einstein-Yang-Mills-dilaton \cite{Stetsko_EYM20} black holes:
\begin{equation}\label{press}
P=-\frac{\Lambda}{16\pi}\left(\frac{b}{r_+}\right)^{2\gamma}.
\end{equation}
Since we have introduced the thermodynamic pressure (\ref{press}), the black hole's mass now should be treated as the thermodynamic function enthalpy $M=H$ \cite{Kastor_CQG09,Kubiznak_CQG17}. Now we can introduce thermodynamic volume for the black hole as a conjugate to the pressure $P$, namely:
\begin{equation}\label{TD_volume}
V=\left(\frac{\partial H}{\partial P}\right)_{S,Q}=\left(\frac{\partial M}{\partial P}\right)_{S,Q}.
\end{equation} 
Having used the upper relation we can write:
\begin{equation}\label{td_vol}
V=\frac{\omega_{n-1}(1+\alpha^2)}{n-\alpha^2}b^{(n-1)\gamma}r_+^{(n-1)(1-\gamma)+1}.
\end{equation} 
Since we have introduced the thermodynamic pressure in the same way as we did for the Eisntein-Maxwell-dilaton \cite{Stetsko_EPJC19} and Einstein-Yang-Mills-dilaton \cite{Stetsko_EYM20} cases it means that the obtained relation for the thermodynamic volume (\ref{td_vol}) takes also the same form as for the mentioned cases. 

It should be also pointed out that the role of the cosmological constant in the framework of AdS/CFT correspondence might be completely different from that we have utilize here. Namely, it might be related to the number of colors $N$ on the field theory side \cite{Kastor_JHEP14,Dolan_JHEP14,Johnson_CQG14,Zhang_JHEP15,Caceres_JHEP15} and consequently the variation of the cosmological constant $\L$ in the bulk gives rise to the change of the number of colors. Since the cosmological constant is related to the number of colors its conjugate quantity might be related to the chemical potential for the colors. It was shown \cite{Karch_JHEP15} that carefully defined holographic dictionary between bulk and boundary theories allows to establish a relation between the first laws of thermodynamics on both sides and derive the so-called holographic Smarr relation form which the Smarr relation in the framework of extended thermodynamics can be derived as a consequence, although there might be some difficulties if corrections beyond large $N$ limit are taken into account or higher order gravity theories are considered. The holographic Smarr relation introduced in \cite{Karch_JHEP15} was extensively studied in various contexts \cite{Caceres_PRD17,Sinamuli_PRD17,Wei_PRD17}. It was also remarked \cite{Karch_JHEP15} that from higher dimensional perspective it is possible to identify the cosmological constant as a pressure in the bulk $P_b$ and corresponding conjugate value as a thermodynamic bulk volume of the black hole $V_b$ and as it was pointed out in \cite{Dolan_JHEP14}, one of the reasons which allows to make this identification is the fact that the cosmological constant $\L$ which appear in solutions derived by dimensional reduction might not have any fundamental role, but it might be on the same footing as the mass or charge of the black hole. 

We also note that there is alternative approach, examined in \cite{Dolan_Entr16} to treat the cosmological constant $\L$ as a thermodynamic variable within the AdS/CFT correspondence. It is based on interpretation of the cosmological constant as the value which gives the length scale in the CFT. Variation of $\L$ gives rise to the variation of the volume in the field theory side on the boundary and as a result a conjugate value to the volume defined by $\L$ would be the pressure, thus the identification of thermodynamic values on the boundary is opposite to their bulk definition. It is worth pointing out that variation of $\L$ might be achieved while the number of colors is held fixed, namely it can performed allowing the Newton's constant $G$ to be varied \cite{Karch_JHEP15}.

Following the work \cite{Corichi_PRD00} we define the global charge of the Yang-Mills field as follows:
\begin{equation}
\bar{Q}=\frac{1}{4\pi\sqrt{(n-1)(n-2)}}\int_{\Sigma}d^{n-1}\chi J(\Omega)\sqrt{Tr(F^{(a)}_{\mu\nu}F^{(a)}_{\mu\nu})}=\frac{\omega_{n-1}}{4\pi}\bar{q}.
\end{equation} 
The latter integral is taken over a sphere which encloses the black hole  and $J(\Omega)$ denotes the Jacobian over angular variables in $n$-dimensional case. From the latter relation we see that the global charge for the Yang-Mills field $\bar{Q}$ is proportional to $\bar{q}$. This charge is also necessary when one tries to obtain the Smarr relation and which will be written below. Now we can also derive the Yang-Mills ``potential'' as a conjugate to the charge written above:
\begin{equation}
\bar{U}=\left(\frac{\partial M}{\partial \bar{Q}}\right)_{S,Q,P}.
\end{equation}
Using the introduced above thermodynamic pressure, Yang-Mills charge and their conjugates we can write the extended first law of black hole thermodynamics:
\begin{equation}\label{ext_first_law}
dM=TdS+VdP+\Phi_e dQ+\bar{U}d\bar{Q}.
\end{equation}
The extended thermodynamics concept also allows us to derive the Smarr relation, which can be written in the form:
\begin{equation}\label{smarr_gen}
(n+\al^2-2)M=(n-1)TS+2(\al^2-1)VP+(n-2+\al\be)\Phi_e Q+(1-\al^2)\bar{U}\bar{Q}.
\end{equation}
In the limit $\al=0$ we arrive at the following relation:
\begin{equation}\label{smarr_simpl}
(n-2)M=(n-1)TS-2VP+(n-2)\Phi_e Q+\bar{U}\bar{Q}.
\end{equation}
One of the central notions in the extended thermodynamics framework is the equation of state for the black hole which is considered as an analog of the Van der Waals equation of state for liquid-gas system. Having used the relations (\ref{temp}) and (\ref{press}) we can represent the equation of state in the form:
\begin{eqnarray}\label{eos}
\nonumber P=\frac{(n-1)}{4(1+\alpha^2)}\frac{T}{r_+}-\frac{(n-1)}{16\pi}\left(\frac{(n-2)}{(1-\alpha^2)}b^{-2\gamma}r_+^{2(\gamma-1)}\times\right.\\\left.\left(1-\bar{q}^2b^{-4\gamma}r^{2(2\gamma-1)}_+\right)+\frac{2q^2}{1-n+\al^2-\al\be}b^{(\k-2(n-1))\g}r^{-2(n-1)(1-\g)-\k\g}_{+}\right).
\end{eqnarray}
To show closer ties of the latter equation with the Van der Waals equation of state one should introduce ``physical'' pressure and temperature instead of geometrical ones that we have in the equation (\ref{eos}). It can be performed as follows \cite{Kubiznak_CQG17}:
\begin{equation}
[P]=\frac{\hbar c}{l^{n-1}_{Pl}}P, \quad [T]=\frac{\hbar c}{k}T,
\end{equation}
 where $l_{Pl}$ is the Planck length for $n+1$--dimensional space-time and $k$ is the Boltzmann constant. We note that after redefinition of the thermodynamic quantities we have a new specific volume in the right hand side of the equation of motion, namely it would be proportional to the product of the horizon radius $r_+$ over $l^{n-1}_{Pl}$. Retaining the geometrical thermodynamic values and introducing new specific ``volume'' we rewrite the equation (\ref{eos}) in the form:
 \begin{eqnarray}\label{eos_2}
\nonumber P=\frac{T}{v}-\frac{(n-1)}{16\pi}\left(\frac{(n-2)}{(1-\alpha^2)}b^{-2\gamma}\bar{\k}^{2(\g-1)}v^{2(\gamma-1)}\left(1-\bar{q}^2b^{-4\gamma}\bar{\k}^{2(2\g-1)}v^{2(2\gamma-1)}\right)\right.\\\left.+\frac{2q^2}{1-n+\al^2-\al\be}b^{(\k-2(n-1))\g}\bar{\k}^{-2(n-1)(1-\g)-\k\g}v^{-2(n-1)(1-\g)-\k\g}\right),
\end{eqnarray}
 where $v$ denotes this specific ``volume'' defined as follows:
\begin{equation}\label{sp_vol}
v=\frac{4(1+\alpha^2)}{n-1}r_+,
\end{equation}
and in the equation of state $\bar{\k}=(n-1)/(4(1+\alpha^2))$. The rewritten equation of state (\ref{eos_2}) is treated as the analog of the  Van der Waals equation of state and can be investigated in similar way. Namely, it allows to examine critical behaviour of the black hole and describe the thermodynamic behaviour of the so called large and small black holes. To investigate critical behaviour first of all we find an inflection point which is defined in completely the same manner as it is done for the standard Van der Waals systems:
\begin{equation}
\left(\frac{\partial P}{\partial v}\right)_T=0, \quad \left(\frac{\partial^2 P}{\partial v^2}\right)_T=0.
\end{equation}  
Utilizing the latter relations we obtain the equation for the critical volume $v_c$:
\begin{equation}\label{cr_vol}
1+3(\al^2-2)\bar{q}^2b^{-4\g}(\bar{\k}v_{c})^{2(2\g-1)}-\frac{2(n-1+\al\be)(3-2n+\al^2-2\al\be)}{(n-2)(1-n+\al^2-\al\be)}q^2b^{(\k-2(n-2))\g}(\bar{\k}v_c)^{-2(n-2)(1-\g)-\k\g}=0.
\end{equation}
The latter equation cannot be solved exactly in general case, here we might find just numerical solutions. We also note that  for $q=0$ the latter equation is reduced to the corresponding equation for the Einstein-Yang-Mills-dilaton case \cite{Stetsko_EYM20} and if $\bar{q}=0$ and $\be=\al$ then we arrive at the relation obtained earlier for the Einstein-Maxwell-dilaton theory \cite{Stetsko_EPJC19}. Now the critical temperature can be represented in the form:
\begin{equation}
T_c=\frac{(n-2)b^{-2\g}\bar{\k}^{2\g-1}v^{2\g-1}_{c}}{\pi(1-\al^2)(2n-3+2\al\be-\al^2)}\left(n+\al\be-2+(\al^2-2)(n-3+\al^2+\al\be)\bar{q}^2b^{-4\g}\bar{\k}^{2(2\g-1)}v^{2(2\g-1)}_{c}\right).
\end{equation}
Finally we can derive the critical pressure:
\begin{equation}
P_c=\frac{(n-2)b^{-2\g}\bar{\k}^{2\g-1}v^{2(\g-1)}_{c}}{4\pi(n-1+\al\be)}\left(n+\al\be-2-3(n+\al^2+\al\be-3)\bar{q}^2b^{-4\g}\bar{\k}^{2(2\g-1)}v^{2(2\g-1)}_{c}\right).
\end{equation} 
Using the obtained relations for the critical temperature $T_c$ and pressure $P_c$ we can write the critical ratio in the form:
\begin{equation}\label{cr_rat}
\rho_c=\frac{P_cv_c}{T_c}=\frac{(1-\al^2)(2n-3+2\al\be-\al^2)\left(n+\al\be-2-3(n+\al^2+\al\be-3)\bar{q}^2b^{-4\g}\bar{\k}^{2(2\g-1)}v^{2(2\g-1)}_{c}\right)}{4(n-1+\al\be)\left(n+\al\be-2+(\al^2-2)(n-3+\al^2+\al\be)\bar{q}^2b^{-4\g}\bar{\k}^{2(2\g-1)}v^{2(2\g-1)}_{c}\right)}.
\end{equation}
The critical ratio (\ref{cr_rat}) shows the dependence not only on the fixed parameters such the dimension $n$ and the coupling constants $\al$, $\be$, but it also depends on the critical volume $v_c$ and the Yang-Mills field parameter $\bar{q}$. It is known that for the Van der Waals system the critical ratio $\rho_c$ has universal character, namely it depends only on the dimension, similar situation was shown to take place for the Reissner-Nordstrom black hole in the framework of extended thermodynamics \cite{Kubiznak_CQG17}. For the Einstein-Maxwell-dilaton black hole the critical ratio depends on the dimension of space $n$ and the coupling parameter $\al$ \cite{Stetsko_EPJC19} and for the Einstein-Yang-Mills-black hole there is the dependence on the coupling parameter $\al$ only \cite{Stetsko_EYM20}. It can be shown easily that in the limit $q=0$ or $\bar{q}=0$ the critical values $v_c$, $T_c$ and $P_c$ are reduced to the corresponding critical values for the Einstein-Yang-Mills-dilaton and Einstein-Maxwell-dilaton cases respectively. We can speculate that here the critical ratio $\rho_c$ might also have universal character, but due to more complicated relation for the critical volume $v_c$ (\ref{cr_vol}) it is not possible to show so easily as it was done before.

\begin{figure}
\centerline{\includegraphics[scale=0.33,clip]{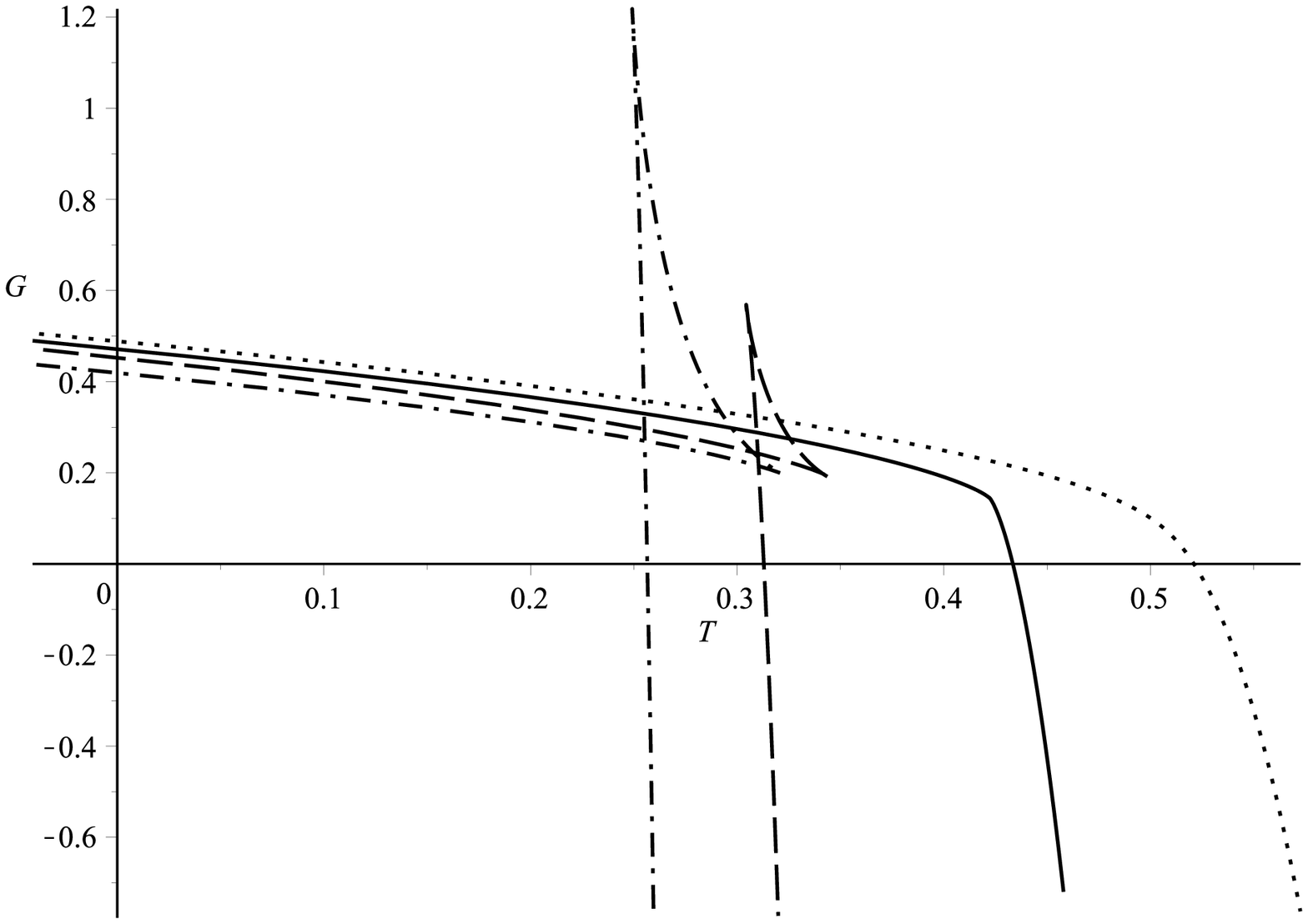}\includegraphics[scale=0.33,clip]{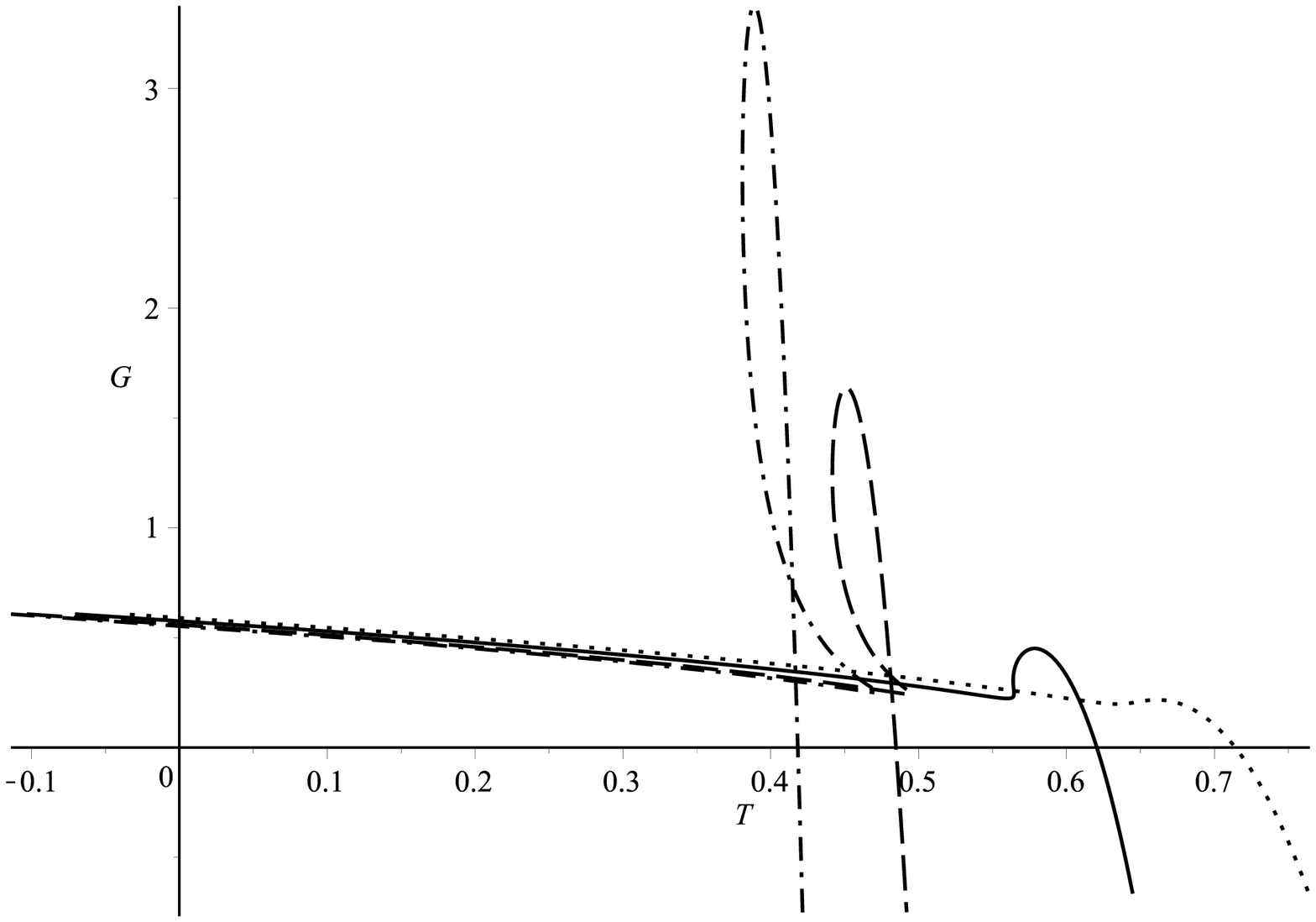}}
\caption{Gibbs free energy $G$ as a function of temperature for fixed values of pressure. For both graphs we have $n=5$, $b=1$, $\be=0.15$, $\bar{q}=0.2$, $q=0.3$ and correspondence of the lines is the following: the dotted, solid, dashed and dash-dotted lines correspond to the pressures $P=3/2P_c$, $P=P_c$, $P=P_c/2$ and $P=P_c/3$ respectively and here $P_c$ denotes the critical pressure, which is defined for given values of $n$, $\al$, $\be$, $b$, $\bar{q}$ and $q$. We also note that for the left graph we have taken $\al=0.1$ and for the right one $\al=0.5$.}\label{gibbs_gr}
\end{figure}

To investigate the critical behaviour of the system better we use the Gibbs free energy, which can be obtained as a Legendre transformation of the enthalpy:
\begin{eqnarray}\label{Gibbs_en}
\nonumber G(T,P)=\frac{\omega_{n-1}(1+\alpha^2)b^{(n-1)\gamma}}{16\pi}r_+^{(n-1)(1-\gamma)}\left(\frac{n-2}{\alpha^2+n-2}b^{-2\gamma}r_+^{2\gamma-1}-\frac{16\pi(1-\alpha^2)P}{(n-1)(n-\alpha^2)}r_{+}\right.\\\left.-\frac{3(n-2)}{(n+3\alpha^2-4)}\bar{q}^2b^{-6\gamma}r_{+}^{3(2\gamma-1)}+\frac{2(2n-3-\al^2+2\al\be)q^2}{(2-n+\al^2-2\al\be)(1-n+\al^2-\al\be)}b^{(\k-2(n-1))\g}r^{1-2(n-1)(1-\g)-\k\g}_{+}\right).
\end{eqnarray}
We note that the latter relation is reduced to corresponding relations for the Gibbs free energy for the Einstein-Maxwell-dilaton black hole if $\bar{q}=0$ \cite{Stetsko_EPJC19} and the Einstein-Yang-Mills-dilaton case if $q=0$ \cite{Stetsko_EYM20}. The expression (\ref{Gibbs_en}) has relatively complicated structure and to analyze analytical behaviour of the Gibbs free energy as a function of the temperature $T$ and the pressure $P$ is a very difficult task, thus to make this behaviour more transparent we give  graphical representation (Figure [\ref{gibbs_gr}]) of the function $G=G(T)$ for several fixed values of the pressure. Both of the graphs on the Figure [\ref{gibbs_gr}] give the $G=G(T)$ dependence for the critical value of pressure $P_c$, for the pressure above critical (here we take $3P_c/2$) and two of the curves correspond to the pressure below the critical one ($P_c/2$, $P_c/3$). In both cases for the pressures above the critical one the Gibbs free energy shows monotonous behaviour, as it is expected. For the critical pressure the Gibbs free energy on the left graph, which corresponds to a smaller value of the parameter $\al$, demonstrates piecewise smooth behaviour, whereas on the right graph it has a specific maximum point which might tell us about a bit different critical behaviour. For the pressures below the critical one the Gibbs free energy on the left graph demonstrates the so-called swallow-tail behaviour which is typical for Van der Waals systems and also for charged black holes \cite{Kubiznak_CQG17}, but in the presence of the dilaton field it takes place just for a relatively small dilaton coupling parameter \cite{Dehyadegari_PRD17,Stetsko_EPJC19,Stetsko_EYM20} and here it is worth noting that the swallow-tail behaviour tells us about the existence of the first order phase transition. On the right graph the Gibbs free energy for the pressures below the critical one in addition to the swallow-tail there is specific domain with a closed loop, as it is known from the cases of the Einstein-Maxwell-dilaton \cite{Dehyadegari_PRD17,Stetsko_EPJC19} and Einstein-Yang-Mills-dilaton theories \cite{Stetsko_EYM20}, here we have a domain not only with the phase transition of the first order, but also there is a domain where the zeroth order phase transition takes place. We also point out here that the domain where the zeroth order phase transition takes place can be investigated similarly as it was done for the Einstein-Maxwell-dilaton \cite{Stetsko_EPJC19} or Einstein-Yang-Mills-dilaton black holes \cite{Stetsko_EYM20}, but from the qualitative side of view the situation here would be almost the same as for the mentioned particular cases.

Similarly as it was performed in the case Einstein-Maxwell-dilaton and Einstein-Yang-Mills-dilaton black holes we can rewrite the equation of state (\ref{eos_2}) in the form:
\begin{equation}\label{red_eos}
p=\frac{1}{\rho_c}\frac{\tau}{\nu}+h(\nu),
\end{equation}
where $p=P/P_c$, $\tau=T/T_c$ and $\nu=v/v_c$ are the so-called reduced variables, $\rho_c$ is the critical ratio (\ref{cr_rat}) and $h(\nu)$ is the function of the reduced volume the evident form of it can be written. The reduced form of the equation of state (\ref{red_eos}) is useful for analysis of thermodynamic behaviour of the system near the critical point, in particular it can be utilized to derive the critical exponents.  We point out here that the evident form the equation (\ref{red_eos}) is completely the same as for the corresponding equation for the Einstein-Maxwell-dilaton \cite{Stetsko_EPJC19} and Einstein-Yang-Mills-dilaton \cite{Stetsko_EYM20} black holes and it means that the critical exponents, in particular $\bar{\beta}$, $\bar{\g}$ and $\bar{\delta}$ are the same as for these previously considered cases, namely we have: $\bar{\beta}=1/2$, $\bar{\g}=1$ and $\bar{\delta}=3$. The critical exponent $\bar{\al}$ which is not derived by virtue of the equation of state (\ref{red_eos}), but is based on the temperature dependence of the entropy is also equal to zero ($\bar{\al}=0$) similarly as it is in the mentioned particular cases \cite{Stetsko_EPJC19,Stetsko_EYM20}, since the entropy (\ref{entropy}) does not depend on the temperature at all, namely it can be shown easily that the entropy would be a function of thermodynamic volume $V$ (\ref{td_vol}) or the specific ``volume'' $v$ (\ref{sp_vol}).  We also point out that under assumptions about existence of the van der Waals critical point and the specific behaviour of the Helmholtz free energy near the critical point it was shown that critical exponents would be the same \cite{Majhi_PLB17}.
\section{Conclusions}
In this paper we have obtained a static spherically symmetric black hole in the framework of the Einstein-Maxwell-Yang-Mills-dilaton theory. This solution can be treated as a generalization of the previously considered Einstein-Yang-Mills-dilaton black hole \cite{Stetsko_EYM20} and from the other side, thus not exactly, of the Einstein-Maxwell-dilaton solution \cite{Stetsko_EPJC19}. The important ingredient of our theory is the dilaton potential of the so-called Liouville form (\ref{dil_pot}) which allowed us to obtain the exact solution (\ref{metric_W}). The parameters (\ref{param_1}) and (\ref{param_2}) of the dilaton potential are chosen to satisfy the equations of motion (\ref{einstein})-(\ref{scal_eq}). It is worth noting that one of the parameters in the dilaton potential (\ref{dil_pot}), namely the parameter $\L$ is not constrained and it can be treated as an effective cosmological constant. In contrast to the Einstein-Maxwell-dilaton case, where Liouville-type dilaton potential $V(\Phi)$ was used in numerous publications, it has not attained so wide interest in case of Einstein-Yang-Mills-dilaton theory, here we point out earlier publication \cite{Radu_CQG05} and our recent work \cite{Stetsko_EYM20}. To define Yang-Mills potential we have used the so-called magnetic Wu-Yang ansatz (\ref{gauge_pot}) which was successfully utilized in numerous papers, where Yang-Mills field was taken into account \cite{Yasskin_PRD75,Mazhari_PRD07,Mazhari_PRD08,Bostani_MPLA10,Dehghani_IJMPD10}.  
Since we consider the static solution, the electromagnetic field potential is chosen in a simple form to give electrostatic field solution (\ref{EM_field}).  The obtained metric function (\ref{metric_W}) has rather complicated structure, but some important features of this function stem directly from the evident form (\ref{metric_W}). Firstly, it should be pointed out that at the infinity the metric $W(r)$ is not asymptotically flat, but also it is not of de Sitter or anti-de Sitter type and this behaviour is mainly defined by the presence of dilaton potential $V(\Phi)$, here we note that similar behaviour at the infinity takes place for Einstein-Maxwell-dilaton \cite{Dehyadegari_PRD17,Stetsko_EPJC19} and Einstein-Yang-Mills-dilaton \cite{Stetsko_EYM20} cases. At the origin the metric function $W(r)$ has singular behaviour, and this singularity is defined by several terms, but the leading term here is caused by the Maxwell-field term. To make the behaviour of the metric $W(r)$ more transparent we give its graphical representation for various values of the parameters $\al$ and $\L$, namely the Figure [\ref{metric_graph}] demonstrates this behaviour. From the Figure [\ref{metric_graph}] we can conclude that the black hole might have two horizons, namely the inner and the outer or event horizons, what is typical for a charged black hole. We also note that the increase of the electric charge $q$ of the black hole while the other parameters are held fixed firstly leads to the decrease of the distance between the horizons, with their following merging into one point, and at that moment the black hole becomes degenerate (extreme), the further increasing of the charge $q$ gives rise to the appearance of a naked singularity. The other important conclusion which can be extracted from the Figure [\ref{metric_graph}] is the fact that the variation of the parameter $\al$ considerably affects on the behaviour of the function $W(r)$ for all distances, while the variation of the other parameters might be substantial in a smaller domain, for instance the change of the parameter $\L$ affects substantially just for large distances (at the asymptotic infinity).

The upper analysis demonstrates that the black hole's metric is ill-defined at several points, namely at the horizon points, at the infinity and at the origin. To analyze whether these points are the points of true physical singularities we have calculated the Kretschmann scalar (\ref{Kr_scal}). The Kretschmann scalar is shown to be finite at the horizon, goes to zero at the infinity and is singular at the origin, and this fact tells us that the only point of physical singularity is the origin what is common for lots of different types of the black holes. We also note that when $\al\rightarrow 0$ the Kretschmann scalar at the infinity tends to a finite value which is of the exactly the same form as for anti-de Sitter black as it should be, because the metric function $W(r)$ turns to be of an anti-de Sitter form at the infinity.

We have also studied the thermodynamics of the black hole, namely we have calculated the temperature (\ref{temp}), entropy (\ref{entropy}) and using the first law of black hole's thermodynamics we derived the mass of the black hole. The analysis of the temperature shows that for large radii of the event horizon the leading term in the temperature (\ref{temp}) is caused by the term related to the cosmological constant $\L$ and this situation is natural, since for other black holes where the cosmological constant is taken into account the corresponding term is dominating at large radii of the horizon. For small horizon radii the leading term in the temperature is caused by the electromagnetic field contribution similarly as it is for the Einstein-Maxwell-dilaton black hole \cite{Stetsko_EPJC19}. For some intermediate values of the horizon radius the temperature might have nonmonotonous behaviour for relatively small $\L$. This nonmonotonicity give us a hint about some critical behaviour. In addition, we have examined the heat capacity (\ref{heat_capac}) $C_{Q}$ which is necessary to analyze thermal stability. The heat capacity was shown to have two discontinuity points and it means that here we have the phase transitions of the Hawking-Page type. It was also shown that for the cosmological constant large enough in its absolute value the discontinuity points might disappear and as a result the the black hole becomes stable for all allowed values of the radius of the horizon. We also note here that in general thermal behaviour of the black hole is very similar to corresponding behaviour in Einstein-Yang-Mills-dilaton \cite{Stetsko_EYM20} and Einstein-Maxwell-dilaton cases \cite{Stetsko_EPJC19}. 

In addition we have studied thermodynamics of the black hole in the framework of the extended thermodynamic phase space. The extended thermodynamics is a very fruitful concept which allowed to find a lot of ties between gravitational and condensed matter physics \cite{Kubiznak_CQG17}. This technique allowed us to derive the equation of state for the black hole (\ref{eos}) and its careful analysis shows some similarity with the well-known Van der Waals equation of state which describes ordinary liquid-gas systems. We have also calculated  the Gibbs free energy of the black hole (\ref{Gibbs_en}). It has been demonstrated that the Gibbs free energy, as a function of the temperature $G=G(T)$ while the pressure $P$ is held fixed has the so-called swallow-tail behaviour for the pressures below the critical one and in case of condensed matter theory it means that there is a phase transition of the first order in this domain. But for our system the situation is a bit more subtle, in addition to the domain where the first order phase transition takes place, here we have the domain where the zeroth order phase transition happens, since there is a range of the pressures below the the critical one, where the Gibbs free energy is discontinuous. The existence of the zeroth order phase transition was recently shown to appear also for other types of the black holes with dilaton fields \cite{Dehyadegari_PRD17,Dayyani_2017,Stetsko_EPJC19,Stetsko_EYM20}. The equation of state as it is shown can be rewritten in the so-called reduced form which is useful for the analysis of the thermal behaviour near the critical point, namely this reduced form allows to derive critical exponents. The reduced equation of state (\ref{red_eos}) takes very similar form as for the Einstein-Maxwell-dilaton \cite{Stetsko_EPJC19} or Einstein-Yang-Mills-dilaton \cite{Stetsko_EYM20} cases, or even as in the case of a charged black hole without the dilaton field  \cite{Kubiznak_JHEP12} and allowed us to make the conclusion that the critical exponents take completely the same values as in these simpler cases.

\section{Acknowledgments}
This work was partly supported by Project FF-83F (No. 0119U002203) from the Ministry of Education and Science of Ukraine.  

\end{document}